# Most productive scale size of China's regional R&D value chain: A mixed structure network

Saeed Assani[1,2]* · Jianlin Jiang[1] · Ahmad Assani[3] · Feng Yang[2]

**Abstract.** This paper offers new mathematical models to measure the most productive scale size (MPSS) of production systems with mixed structure networks (mixed of series and parallel). In the first property, we deal with a general multi-stage network which can be transformed, using dummy processes, into a series of parallel networks. In the second property, we consider a direct network combined with series and parallel structure. In this paper, we propose new models to measure the overall MPSS of the production systems and their internal processes. MPSS decomposition is discussed and examined. As a real-life application, this study measures the efficiency and MPSS of research and development (R&D) activities of Chinese provinces within an R&D value chain network. In the R&D value chain, profitability and marketability stages are connected in series, where the profitability stage is composed of operation and R&D efforts connected in parallel. The MPSS network model provides not only the MPSS measurement but also values that indicate the appropriate degree of intermediate measures for the two stages. Improvement's strategy is given for each region based on the gap between the current and the appropriate level of intermediate measures. Our findings show that the marketability efficiency values of Chinese R&D regions were low, and no regions are operated under the MPSS. As a result, most Chinese regions performed inefficiently regarding both profitability and marketability. This finding provides initial evidence that the generally lower profitability and marketability efficiency of Chinese regions is a severe problem that may be due to wasted resources

* Saeed Assani
  saeedassani@nuaa.edu.cn
  Tel: +8615077820900
  Jianlin Jiang
  jiangjianlin@nuaa.edu.cn
  Ahmad Assani
  ahmad.assani@hs-karlsruhe.de
  Feng Yang
  fengyang@ustc.edu.cn
[1] College of Science, Nanjing University of Aeronautics and Astronautics, Nanjing 210016, China
[2] School of Management, University of Science and Technology of China, Hefei 230026, China
[3] Faculty of Computer Science and Business Computer Systems, Karlsruhe University of Applied Science, Karlsruhe, 76133, Germany

on production and R&D.

# 1 Introduction

Standard data envelopment analysis (DEA), proposed by (Charnes, Cooper, & Rhodes, 1978), treats the decision making units (DMUs) as a black box. When we open this box, exciting findings and results can be obtained. The decision makers can see precisely the source of inefficiency in their systems and thus, network DEA is conducted. Kao (2014) reviewed and classified the studies on network DEA by examining the models used and the structures of the network systems of the problem being studied. In his classification, several structures of network DEA have been discussed. The most used structures are series, parallel, mixed, and dynamic.

As we mentioned above, one of the known structures is called a mixed structure, which is neither series nor parallel, but a mixture of them. In general, the mixed structure network is a little bit complex than series or parallel structures. One of the famous techniques to deal with mixed structure networks is using dummy processes to transform the original mixed structure network to a more informal network, series, or parallel. More specifically, the key to evaluating the system efficiency of such a network is to find a transformation into the underlying structures, series, or parallel.

In DEA literature, mixed structure networks are applied in different areas of efficiency evaluation. Adler, Liebert, & Yazhemsky (2013) evaluated the performance of European airports using a mixed structure network, two stages of operations. The first stage generates passengers and cargo, and the second stage is composed of two processes of aeronautical and non-aeronautical activities. (Yu, 2010) also evaluated the airport performance by decomposing the operations intro services and production, where the former were further divided into landside and airside in parallel. (Lin & Chiu, 2013) divided the bank's operation systems into three stages, profitability, services, and production, where services were further decomposed into consumer and corporate banking processes. (Hsieh & Lin, 2010) evaluated the hotel's efficiency using two stages, production and service, where the production stage is further separated into rooms and restaurants in parallel.

The assessment of regional research and development (R&D) activities is an essential task in promoting and maintaining the development of scientific and technological (S&T) investment and management in a regional economy. This topic has received increasing academic interest in recent years (Del Monte & Papagni, 2003; Hagedoorn & Cloodt, 2003; Hu, 2001; Jacobides & Winter, 2005; Liu & Lu, 2010; Porter & Roach, 1996; Roach, 1996). In DEA literature, few studies mentioned R&D efficiency with a mixed structure network. (Wang, Lu, Huang, & Lee, 2013) proposed a mixed structure network of two stages to study the profitability and marketability efficiencies of high-technology firms. The first stage, profitability efficiency, is separated into basic production and R&D efforts in parallel, while the second stage is the marketability efficiency. However, these types of studies do not offer enough information on the productivity scale size of the evaluated DMUs. As it is known, an efficient DMU is not necessary to be MPSS. Therefore, it is essential to know the scale size of the evaluated Chinese regions and select those regions that achieve the most productive scale size. In other words, there is a need to describe the relationship among the most productivity scale size (MPSS) of the different processes in a mixed structure network. For example, how does the MPSS of R&D efforts affect the overall MPSS of the R&D value chain? How much wasted resources, in the inputs or in the intermediate measures between the profitability and marketability stages, can be allocated and optimally invested? This leads to these research questions: (i) How is the MPSS of the overall and internal processes can be estimated? (ii) How is the relationship between the overall MPSS and the internal processes is derived?

When we could identify the internal processes, which do not achieve the MPSS state, the question is (iii) How the non-MPSS decision making units (DMUs) can be moved to the MPSS region to achieve the best economic scale?

This study contributes to the methodological and applications level. At the methodological level, new MPSS models are introduced to deal with two properties of mixed structure networks. The first property is a general multi-stage system where each stage has its exogenous inputs and produces two types of outputs, intermediated measures that enter the next stage and the final output. The second property is a classical mixed of series-parallel networks. Beyond the theoretical content, this study reports an application of China's regional R&D value chain network. The application's network structure is the same as the R&D value chain described in (Wang et al., 2013), but it is applied to the Chinese regions instead of high-technology firms. One difference is that we are using more inputs in the R&D process than in (Wang et al., 2013). The introduced R&D value chain network has a two-stage structure (profitability and marketability stages). The profitability stage has two processes, operational and R&D efforts, connected in parallel. In the proposed application, we are aiming to measure the efficiency and the most productive scale size of China's R&D value chain. The efficiencies and MPSSs of the operational and R&D efforts processes and marketability stage are measured from 2014 to 2015. Improvement's strategy is given for each Chinese region based on the gap between the current and the appropriate levels of intermediate measures that connect the profitability and marketability stages.

This study is organized as follows. The second section deals with the first property; MPSS for a general multi-stage system. The third section introduces China's regional R&D value chain network. Discussion and results are displayed in Section 4. Section 5 concludes the study.

## 2 MPSS for a general multi-stage system

In this section, we consider a general multi-stage system where each stage has its own exogenous inputs and produce two kinds of outputs. The first output is the final output of this stage, and the second output is an intermediate measure, which further will enter the next stage as input. Since each stage has its own exogenous input, this network can be seen as a parallel network. The difference between this network and the classical parallel network is that the latter does not consider the intermediate measures that may arise from one stage to another. Thus, we can say that the evaluation of such a network is a complicated task.

In network DEA literature, one of the known approaches to deal with the general multi-stage network is to look for a transformation to a typical network such as series, parallel, or combination of them. This transformation can be done using dummy processes. These dummy processes will enter the system as efficient processes and help to transform the system into a tandem system. The transformed system will be more accessible to be interpreted and evaluated.

The popular non-life insurance industry has a network structure. In fact, it has two processes in its operation, the insurance service itself and capital investment. This problem has been studied many times in the literature (Kao & Hwang, 2008) as a two-stage problem, in which insurance service is the first stage, and capital investment is the second. The inputs of the considered system are insurance expenses (X1) and investment expenses (X2). There are two types of intermediate products, direct written premiums (Z1) and reinsurance premiums (Z2). The outputs of the system are underwriting profit (Y1) and investment profit (Y2).

To make this application consistent with the multi-stage network case, we associate the investment expenses (X2) with the capital investment process rather than the insurance service process, and the underwriting profit (Y1) is the profit generated from the insurance service process instead of the capital investment process. In this sense, the system is not a simple series system, but a network system as depicted in Figure 1, where the insurance service process uses insurance expenses (X1) to produce underwriting profit (Y1), direct written premiums (Z1), and reinsurance premiums (Z2). The capital investment process uses investment expenses (X2), direct written premiums (Z1), and reinsurance premiums (Z2) to produce investment profit (Y2).

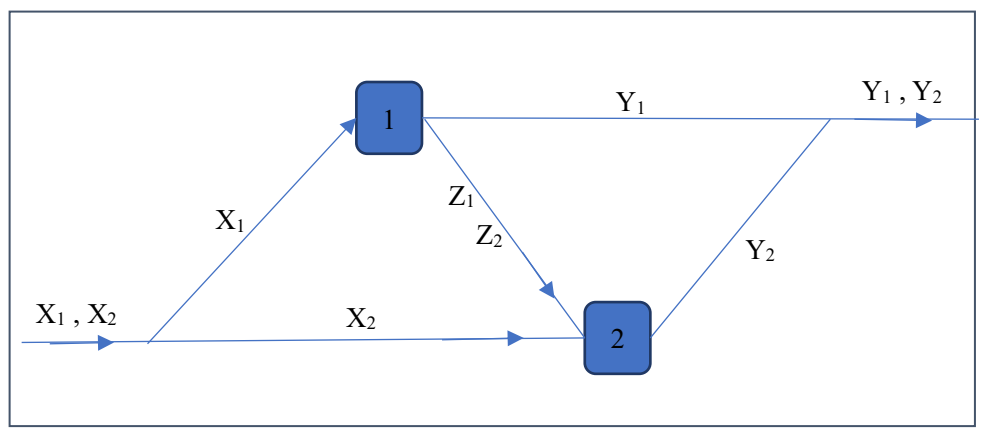

**Figure 1 Network structure of the non-life insurance operation system**

In the next, we introduce the MPSS model for the network system in Figure 1.

## 2.1 The proposed MPSS models for a general multi-stage system

Following the concept of the MPSS model discussed in (Assani, Jiang, Cao, & Yang, 2018), the MPSS model of the general multi-stage network displayed in Figure

1 can be derived as follows.

$$MPSS^{S*} = \text{Max} \quad \theta_2^2 - \theta_1^1 \tag{1}$$

$$s.t. \sum_{j=1}^{n} \lambda_j^1 X_{ij}^1 \leq \theta_1^1 X_{io}^1, i = 1,2,\ldots,m^{(1)}$$

$$\sum_{j=1}^{n} \lambda_j^1 Z_{dj}^1 \geq \tilde{Z}_d^1, d = 1,2,\ldots,p^{(1)}$$

$$\sum_{j=1}^{n} \lambda_j^1 Z_{dj}^2 \geq \tilde{Z}_{do}^2, d = 1,2,\ldots,p^{(2)}$$

$$\sum_{j=1}^{n} \lambda_j^1 Y_{rj}^1 \geq \theta_2^1 Y_{ro}^1, r = 1,2,\ldots,s^{(1)}$$

$$\sum_{j=1}^{n} \lambda_j^2 Z_{dj}^1 \leq \tilde{Z}_{do}^1, d = 1,2,\ldots,p^{(1)}$$

$$\sum_{j=1}^{n} \lambda_j^2 Z_{dj}^2 \leq \tilde{Z}_{do}^2, d = 1,2,\ldots,p^{(2)}$$

$$\sum_{j=1}^{n} \lambda_j^2 X_{ij}^2 \leq \theta_1^2 X_{io}^2, i = 1,2,\ldots,m^{(2)}$$

$$\sum_{j=1}^{n} \lambda_j^2 Y_{rj}^2 \geq \theta_2^2 Y_{ro}^2, r = 1,2,\ldots,s^{(2)}$$

$$\sum_{j=1}^{n} \lambda_j^t = 1, t = 1,2$$

$$\theta_1^1, \theta_2^h, \lambda_j^t \geq 0, j = 1,2,\ldots,n, t = 1,2,$$

where $\theta_1^1$ and $\theta_2^h(h = 1,2)$ are scalars representing expansion or contraction factors applied to the two stages' inputs and outputs of the evaluated DMU. The objective of model (1) is to maximize $\theta_2^2 - \theta_1^1$, which will reduce the inputs of the two stages proportionally (radially) to $\theta_1^1 X_o^1$ and $\theta_1^2 X_o^2$ as small as possible and raise the outputs of the two stages proportionally to $\theta_2^1 Y_{ro}^1$ and $\theta_2^2 Y_{ro}^2$ as large as possible. This model also generates a set of new intermediate measures $\tilde{Z}_{do}^1$ and $\tilde{Z}_{do}^2$, which helps the decision makers to achieve the most productive scale size.

From one point of view, the main difference between the MPSS in the black box and the general multi-stage DEA is that the latter considers the procedures are taking place inside the evaluated DMU while the former does not. More specifically, new intermediate measures are generated in model (1), but they are ignored in the black-box MPSS model. This leads us to expect that the general multi-stage MPSS model is more discriminative than the black-box MPSS model (Assani et al., 2018).

From another point of view, the difference between the multi-stage DEA network and the general multi-stage DEA network is that the later has exogenous inputs for each internal stage. While the difference between the general multi-stage DEA network and the classical parallel network is that, the former has intermediate measures connecting the internal processes.

Now we define the system MPSS for a general multi-stage network DEA.

**Definition 1** $DMU_o$ is (overall) MPSS if and only if the optimal objective function value of model (1) is zero.

Model (1) generates a new set of intermediate measures, optimal intermediate measures, that help the evaluated DMU to achieve the most productive scale size. Instead of considering the intermediate measures as variables in the MPSS model, another approach is to adjust them radially (proportionally) as the inputs and the outputs of each internal stage. The resulting model will have the following formula.

$$MPSS^{S*} = \text{Max} \quad \theta_2^2 - \theta_1^1 \qquad (2)$$

$$s.t. \sum_{j=1}^{n} \lambda_j^1 X_{ij}^1 \leq \theta_1^1 X_{io}^1, i = 1,2,\ldots,m^{(1)}$$

$$\sum_{j=1}^{n} \lambda_j^1 Z_{dj}^1 \geq \theta_2^1 Z_{do}^1, d = 1,2,\ldots,p^{(1)}$$

$$\sum_{j=1}^{n} \lambda_j^1 Z_{dj}^2 \geq \theta_2^1 Z_{do}^2, d = 1,2,\ldots,p^{(2)}$$

$$\sum_{j=1}^{n} \lambda_j^1 Y_{rj}^1 \geq \theta_2^1 Y_{ro}^1, r = 1,2,\ldots,s^{(1)}$$

$$\sum_{j=1}^{n} \lambda_j^2 Z_{dj}^1 \leq \theta_1^2 Z_{do}^1, d = 1,2,\ldots,p^{(1)}$$

$$\sum_{j=1}^{n} \lambda_j^2 Z_{dj}^2 \leq \theta_1^2 Z_{do}^2, d = 1,2,\ldots,p^{(2)}$$

$$\sum_{j=1}^{n} \lambda_j^2 X_{ij}^2 \leq \theta_1^2 X_{io}^2, i = 1,2,\ldots,m^{(2)}$$

$$\sum_{j=1}^{n} \lambda_j^2 Y_{rj}^2 \geq \theta_2^2 Y_{ro}^2, r = 1,2,\ldots,s^{(2)}$$

$$\sum_{j=1}^{n} \lambda_j^t = 1, t = 1,2$$

$$\theta_1^1, \theta_2^h, \lambda_j^t \geq 0, j = 1,2,\ldots,n, \ t = 1,2.$$

Since $Z^1, Z^2$, and $Y^1$ are the outputs of the first stage, they have the same distance measure $\theta_2^1$. Similarly, $X^2, Z^1$, and $Z^2$ have the same distance measure $\theta_1^2$.

Both models (1) and (2) have the same objective function that maximizes the productivity average of the inputs and the outputs of the whole system but in different ways. More specifically, the difference between models (1) and (2) is that the former looks for the optimal intermediate measures that connect the internal stages in order to achieve the most productive scale size, while the latter adjusts the intermediate measures in proportional scale as applied to the inputs and outputs in the standard MPSS model.

To obtain the MPSS of the internal stages, we adopt models (1) and (2) with two simple modifications. The first is to replace the objective function to be $\theta_2^1 - \theta_1^1$ and $\theta_2^2 - \theta_1^2$ for stage 1 and stage 2, respectively. The second is to maintain the MPSS value of the system while measuring the MPSS for the first stage and maintain both system MPSS and stage 1 MPSS values while measuring the MPSS of stage 2.

The MPSS of stage 1 is given as follows.

$$MPSS^{I*} = \text{Max} \quad \theta_2^1 - \theta_1^1 \tag{3}$$

$$s.t. \sum_{j=1}^{n} \lambda_j^1 X_{ij}^1 \leq \theta_1^1 X_{io}^1, i = 1,2,\ldots,m^{(1)}$$

$$\sum_{j=1}^{n} \lambda_j^1 Z_{dj}^1 \geq \theta_2^1 Z_{do}^1, d = 1,2,\ldots,p^{(1)}$$

$$\sum_{j=1}^{n} \lambda_j^1 Z_{dj}^2 \geq \theta_2^1 Z_{do}^2, d = 1,2,\ldots,p^{(2)}$$

$$\sum_{j=1}^{n} \lambda_j^1 Y_{rj}^1 \geq \theta_2^1 Y_{ro}^1, r = 1,2,\ldots,s^{(1)}$$

$$\sum_{j=1}^{n} \lambda_j^2 Z_{dj}^1 \leq \theta_1^2 Z_{do}^1, d = 1,2,\ldots,p^{(1)}$$

$$\sum_{j=1}^{n} \lambda_j^2 Z_{dj}^2 \leq \theta_1^2 Z_{do}^2, d = 1,2,\ldots,p^{(2)}$$

$$\sum_{j=1}^{n} \lambda_j^2 X_{ij}^2 \leq \theta_1^2 X_{io}^2, i = 1,2,\ldots,m^{(2)}$$

$$\sum_{j=1}^{n} \lambda_j^2 Y_{rj}^2 \geq \theta_2^2 Y_{ro}^2, r = 1,2,\ldots,s^{(2)}$$

$$\sum_{j=1}^{n} \lambda_j^t = 1, t = 1,2$$

$$\theta_2^2 - \theta_1^1 = MPSS^{S*}$$

$$\theta_1^1, \theta_2^h, \lambda_j^t \geq 0, j = 1,2,\ldots,n,\ t = 1,2.$$

Similarly, the MPSS model for stage 2 is given as follows.

$$MPSS^{II*} = \text{Max}\quad \theta_2^2 - \theta_1^2 \tag{4}$$

$$s.t. \sum_{j=1}^{n} \lambda_j^1 X_{ij}^1 \leq \theta_1^1 X_{io}^1, i = 1,2,\ldots,m^{(1)}$$

$$\sum_{j=1}^{n} \lambda_j^1 Z_{dj}^1 \geq \theta_2^1 Z_{do}^1, d = 1,2,\ldots,p^{(1)}$$

$$\sum_{j=1}^{n} \lambda_j^1 Z_{dj}^2 \geq \theta_2^1 Z_{do}^2, d = 1,2,\ldots,p^{(2)}$$

$$\sum_{j=1}^{n} \lambda_j^1 Y_{rj}^1 \geq \theta_2^1 Y_{ro}^1, r = 1,2,\ldots,s^{(1)}$$

$$\sum_{j=1}^{n} \lambda_j^2 Z_{dj}^1 \leq \theta_1^2 Z_{do}^1, d = 1,2,\ldots,p^{(1)}$$

$$\sum_{j=1}^{n} \lambda_j^2 Z_{dj}^2 \leq \theta_1^2 Z_{do}^2, d = 1,2,\ldots,p^{(2)}$$

$$\sum_{j=1}^{n} \lambda_j^2 X_{ij}^2 \leq \theta_1^2 X_{io}^2, i = 1,2,\ldots,m^{(2)}$$

$$\sum_{j=1}^{n} \lambda_j^2 Y_{rj}^2 \geq \theta_2^2 Y_{ro}^2, r = 1,2,\ldots,s^{(2)}$$

$$\sum_{j=1}^{n} \lambda_j^t = 1, t = 1,2$$

$$\theta_2^2 - \theta_1^1 = MPSS^{S*}$$

$$\theta_2^1 - \theta_1^1 = MPSS^{I*}$$

$$\theta_1^1, \theta_2^h, \lambda_j^t \geq 0, j = 1,2,\ldots,n, \ t = 1,2.$$

## 2.2 MPSS decomposition

(Kao, 2009) proposed an approach to transform a general multi-stage network system into one of the series and parallel structures. In his approach, the longest path of processes in the system is used as the backbone of the transformed system, and dummy processes are introduced to carry the inputs and outputs of intermediate processes.

A dummy process has the same inputs and outputs, and they are used only to help the representation. The resulting system has two stages connected in series. At each stage, one dummy process, connected in parallel with a real process, is added to carry the inputs to be used in the next stage and the outputs produced in the first stage. Figure 2 shows the transformation of the system in Figure 1, where circles and squares represent the dummy and the real processes, respectively.

The tandem system, the transformed system, has two stages connected in series. Based on (Assani et al., 2018), the tandem system MPSS is the sum of the MPSS values of the two stages.

$$MPSS^{tandem} = MPSS^{stage1} + MPSS^{stage2} \qquad (5)$$

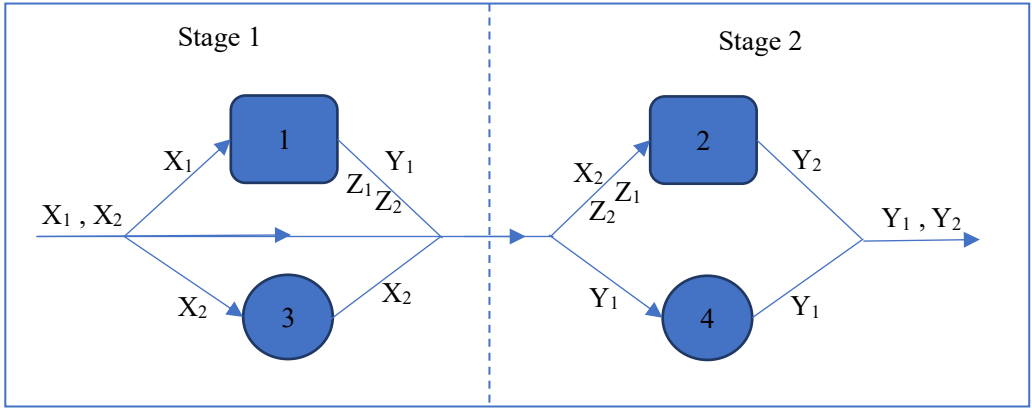

**Figure 2 Equivalent tandem system of the non-life insurance operation system**

Each stage in the tandem system is a classical parallel system. Based on the MPSS decomposition of parallel network systems described in (Assani, Jiang, Assani, & Yang, 2019), the MPSS of each stage is the weighted sum of the MPSS of the real and the dummy processes. Since the dummy process produces the same amount of the consumed inputs, thus, it is the MPSS process, and its MPSS value is zero. The MPSSs of the two stages are given as follows.

$$MPSS^{stage1} = \omega^1.MPSS^I + (1-\omega^1).MPSS^{dummy1} = \omega^1.MPSS^I. \qquad (6)$$

$$MPSS^{stage2} = \omega^2.MPSS^{II} + (1-\omega^2).MPSS^{dummy2} = \omega^2.MPSS^{II}. \qquad (7)$$

Where $\omega^1$ and $\omega^2$ are the importance of process 1 and process 2, respectively, in the classical parallel systems. It is known that the multiplier DEA form has the ability to put a restriction on the weights of the inputs and outputs of the internal stages, and the ability to assume the importance of the internal processes in the parallel network structure. Here in the MPSS concept, our task is deriving the points on the efficient frontier that represent the most productive scale size. This is called the target setting. Therefore, we will not consider any assurance region (AR) in our models. In addition, we will assume that the internal processes of the parallel network have the same relative importance, that is, choosing one process of the parallel network by the DMU has the same importance of choosing the other processes in the network. In the calculation of Table 1, we assume that $\omega^1 = \omega^2 = 0.5$.

Using the previous MPSS decomposition, the tandem system MPSS, the network MPSS, the two processes MPSS, and the two stages MPSS are reported in Table 1.

**Table 1 MPSS measures of the 24 non-life insurance companies**

| DMUs | Black-box MPSS | Tandem system MPSS | Network system MPSS | Process 1 | Process 2 | Stage 1 | Stage 2 |
|---|---|---|---|---|---|---|---|
| 1 | 0.0750 | 0.2039 | 0.4079 | 0.1254 | 0.2825 | 0.0627 | 0.1413 |
| 2 | 0.0000 | 0.2090 | 0.4179 | 0.0000 | 0.4179 | 0.0000 | 0.2090 |
| 3 | 0.0585 | 0.2217 | 0.4434 | 0.4434 | 0.0000 | 0.2217 | 0.0000 |
| 4 | 10.9947 | 2.8362 | 5.6724 | 0.8149 | 4.8575 | 0.4075 | 2.4287 |
| 5 | 0.0000 | 0.0000 | 0.0000 | 0.0000 | 0.0000 | 0.0000 | 0.0000 |
| 6 | 0.7937 | 1.2915 | 2.5830 | 0.7414 | 1.8415 | 0.3707 | 0.9208 |
| 7 | 0.6002 | 2.1819 | 4.3639 | 0.0012 | 4.3627 | 0.0006 | 2.1813 |
| 8 | 0.3554 | 1.6132 | 3.2265 | 0.1665 | 3.0600 | 0.0833 | 1.5300 |
| 9 | 3.2489 | 1.5207 | 3.0414 | 0.0000 | 3.0414 | 0.0000 | 1.5207 |
| 10 | 0.3290 | 0.4405 | 0.8809 | 0.0000 | 0.8809 | 0.0000 | 0.4405 |
| 11 | 1.1975 | 57.3520 | 114.7040 | 0.8877 | 113.816 | 0.4439 | 56.908 |
| 12 | 0.0000 | 0.2350 | 0.4699 | 0.4699 | 0.0000 | 0.2350 | 0.0000 |
| 13 | 0.2267 | 3.7035 | 7.4070 | 0.0000 | 7.4070 | 0.0000 | 3.7035 |
| 14 | 1.1417 | 2.7598 | 5.5195 | 0.2932 | 5.2263 | 0.1466 | 2.6132 |
| 15 | 0.0006 | 0.2776 | 0.5552 | 0.1364 | 0.4187 | 0.0682 | 0.2094 |
| 16 | 2.2001 | 3.6744 | 7.3488 | 0.9137 | 6.4351 | 0.4569 | 3.2175 |
| 17 | 0.0247 | 0.4993 | 0.9987 | 0.0000 | 0.9987 | 0.0000 | 0.4993 |
| 18 | 2.6003 | 1.6090 | 3.2181 | 0.0000 | 3.2181 | 0.0000 | 1.6090 |
| 19 | 0.0308 | 1.1551 | 2.3103 | 0.0000 | 2.3103 | 0.0000 | 1.1551 |
| 20 | 0.0172 | 0.9978 | 1.9955 | 0.0000 | 1.9955 | 0.0000 | 0.9978 |
| 21 | 39.2892 | 78.0128 | 156.0257 | 17.7392 | 138.286 | 8.8696 | 69.143 |
| 22 | 0.0000 | 0.0000 | 0.0000 | 0.0000 | 0.0000 | 0.0000 | 0.0000 |
| 23 | 82.4934 | 15.6996 | 31.3992 | 11.0917 | 20.3075 | 5.5459 | 10.153 |
| 24 | 14.1229 | 1.3540 | 2.7079 | 0.0000 | 2.7079 | 0.0000 | 1.3540 |

The second column of Table 1 reports the black-box MPSS scores calculated based on (Banker, 1984). Of 24 companies, four are MPSS. The network system MPSS scores calculated by model (1) are listed in the fourth column. Two of the previous four MPSS companies are MPSS under the network MPSS model (1). That shows that the network MPSS model is more discriminative than the black-box MPSS model. Another option of network MPSS model is the ability to decompose the MPSS into the processes MPSS as it is reported in the 5$^{th}$ and 6$^{th}$ columns. The two processes MPSS scores show that there are eleven MPSS companies in the first process, while there are four MPSS companies are in the other process.

The tandem system MPSS and the two stages MPSS scores are listed in the 3$^{rd}$, 7$^{th}$, and 8$^{th}$ columns of Table 1. Based on the MPSS decomposition of the series network structure (Assani et al., 2018), the tandem system MPSS is the sum of the two stages MPSSs. Since the dummy process is efficient and produces the same amount that consumes, it is MPSS process, and its MPSS score is zero. Remember that we selected the importance of the real process and the dummy process to be the same. It is evident that the stage 1's MPSS is the weighted sum of the real process 1 and the dummy process 1 which is satisfied with the MPSS parallel decomposition (Assani et al., 2019). The last two columns of Table 1 show the MPSS decomposition of the two stages.

## 3 R&D value chain network

As an application to a mixed structure network, we introduce the R&D value chain of China's regional R&D activities. This chain has been studied before (Wang et al., 2013). In that paper, the authors proposed and verified an R&D value chain framework to explore the relationship between R&D, productivity, and firm market values. The proposed chain is a mixed structure network composed of two stages connected in series, where the first stage has two processes connected in parallel.

Here we reuse this network, with simple modifications, to measure the MPSS of China's regional R&D activities considering the production, R&D efforts, and market value (see Figure 3). We will first report (Wang et al., 2013)'s R&D value chain DEA model. Then we compute the overall, operation, R&D, and marketability efficiencies for the Chinese regions. In the next step, we propose our R&D value chain MPSS model. Then the MPSS of the two stages will be measured.

### 3.1 Specification of input and output variables

We consider the number of employees as first input in the operation process as the employees help the firms to engage in the production process (Becheikh, Landry, & Amara, 2006; Sterlacchini, 1999). One variable is the investment on assets, which included the investments on the standard resources that support R&D innovation activities. These two inputs are the generators in the primary production process and produce the sales volume as an output. Sales volume represents the profitability associated with R&D and innovation activities (Thornhill, 2006).

In the R&D efforts process, R&D personnel is a significant input as well as the R&D projects and the R&D expenditure. These inputs together are aiming to achieve the research targets, especially the patents that are the most critical output in the R&D activities. More specifically, the number of R&D personnel is an essential indicator for motivating firms to become involved in R&D innovation activities (Zhong, Yuan, Li, & Huang, 2011). The R&D expenditures are often considered one of the critical factors when we evaluate the efficiency at the firm level (Griliches, 1998). It is noted that the R&D expenditure is considered as one of the critical indicators that increase the profitability efficiency (Capon, Farley, & Hoenig, 1990), especially for innovation inputs (Graves & Langowitz, 1996; Hitt, Hoskisson, & Kim, 1997; Zhong et al., 2011). R&D projects are the third inputs of the R&D efforts' process. The number of R&D projects reflects the available opportunities for researchers to be creative. In addition, R&D projects are the primary field that the researchers can get their patents.

The output of the R&D efforts' process is the technical knowledge, which can be in the form of patents (Deeds & Decarolis, 1999; Graves & Langowitz, 1996; Hall & Ziedonis, 2001; Hitt et al., 1997).

Together, the primary production and R&D efforts constitute the profitability stage. As a result, sales volume and patents are obtained from the profitability stage. Although these indicators can describe the production and R&D performance of Chinese regions, they do not reflect their market valuations. Therefore, we follow (Seiford & Zhu, 1999) and consider the marketability stage as an additional stage to be incorporated with the profitability stage. In the marketability stage, sales volume and patents are used as inputs and the market value which is the replacement value of its tangible assets (Blundell, Griffith, & Van Reenen, 1999; Seiford & Zhu, 1999), has been selected as the final output of the marketability stage.

In China's R&D value chain network, eight productivity performance indicators were used. In terms of profitability efficiency, the current study employed five inputs: the number of employees, the investments on the assets from production activities, R&D personnel, R&D projects, and R&D expenditures. The two outputs were the sales volume and the number of patents. For marketability efficiency, there are two inputs: sales volume and the number of patents and one output, market value, in the second stage as it is displayed in Figure 3.

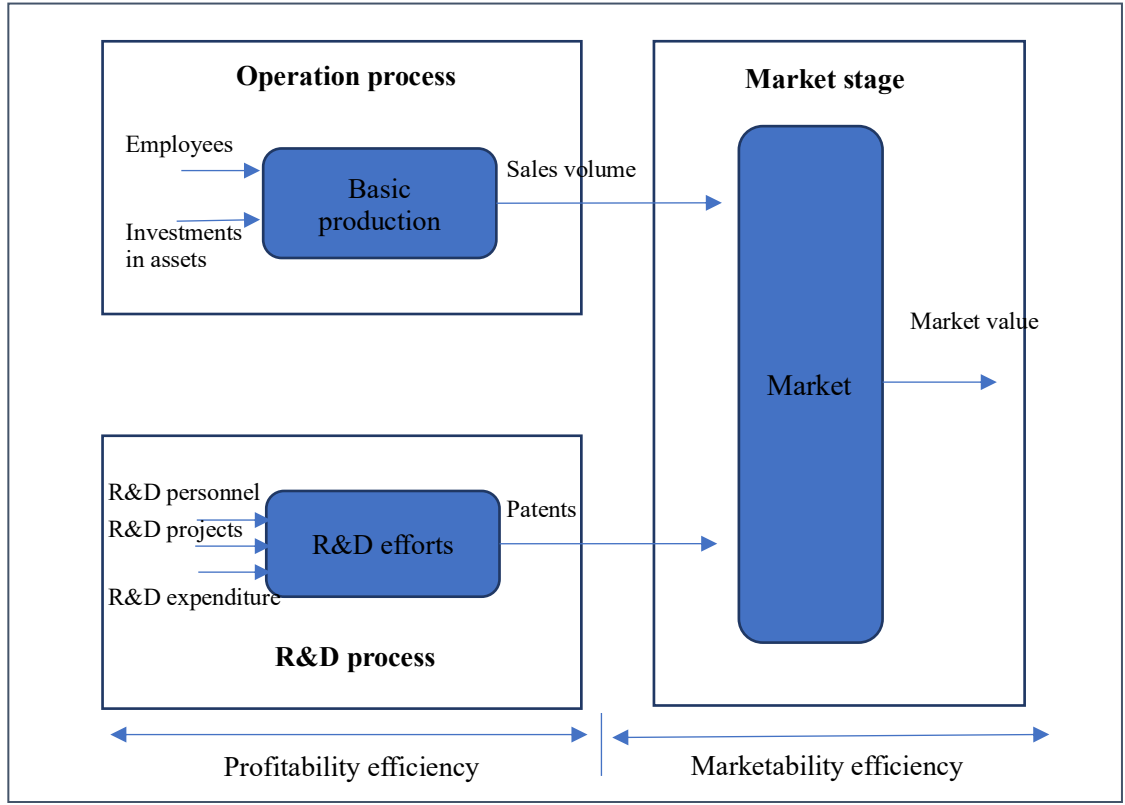

**Figure 1 R&D value chain structure**

## 3.2 Specification of the R&D value chain DEA model

(Wang et al., 2013) proposed a network DEA model to measure the efficiencies of the network displayed in Figure 3. In their model, the overall, operation, R&D, and marketability efficiencies can be computed in one-step as follows:

$$Chain's\ efficiency\ =\ \text{Min}\ \omega_1 \theta^O + \omega_2 \theta^R - \omega_3 \theta^M \quad (8)$$
$$Stage\ 1: Operation$$
$$s.t. \sum_{j=1}^{n} \lambda_j X_{ij}^O \leq \theta^O X_{io}^O, i = 1,2,\dots,m$$
$$\sum_{j=1}^{n} \lambda_j Z_{dj}^O \geq \tilde{Z}_{do}^O, d = 1,2,\dots,p$$
$$\sum_{j=1}^{n} \lambda_j = 1$$
$$\theta^O \leq 1, \lambda_j \geq 0, j = 1,2,\dots,n$$
$$Stage\ 1: R\&D$$
$$\sum_{j=1}^{n} \mu_j X_{kj}^R \leq \theta^R X_{ko}^R, k = 1,2,\dots,K$$

$$\sum_{j=1}^{n} \mu_j Z_{ej}^R \geq \tilde{Z}_{eo}^R, e = 1,2,\dots,E$$

$$\sum_{j=1}^{n} \mu_j = 1$$

$$\theta^R \leq 1, \mu_j \geq 0, j = 1,2,\dots,n$$

*Stage 2 : Market*

$$\sum_{j=1}^{n} \varphi_j Z_{dj}^O \leq \tilde{Z}_{do}^O, d = 1,2,\dots,p$$

$$\sum_{j=1}^{n} \varphi_j Z_{ej}^R \leq \tilde{Z}_{eo}^R, e = 1,2,\dots,E$$

$$\sum_{j=1}^{n} \varphi_j Y_{rj} \geq \theta^M Y_{ro}, r = 1,2,\dots,s$$

$$\sum_{j=1}^{n} \varphi_j = 1$$

$$\theta^M \geq 1, \varphi_j \geq 0, j = 1,2,\dots,n,$$

where $X_{ij}^O$ $(i = 1,2,\dots,m; j = 1,2,\dots,n)$ denotes the inputs of the operations stage and R&D stage that are used to produce the outputs $Z_{dj}^O$ $(d = 1,2,\dots,p; j = 1,2,\dots,n)$. In the same way, the $X_{kj}^R$ $(k = 1,2,\dots,K; j = 1,2,\dots,n)$ are the inputs of the R&D stage that are used to produce the outputs represented by $Z_{ej}^R$ $(e = 1,2,\dots,E; j = 1,2,\dots,n)$. Then $Z_{dj}^O$ and $Z_{ej}^R$ are employed as inputs in the second stage to produce the final outputs $Y_{rj}$ $(r = 1,2,\dots,s; j = 1,2,\dots,n)$.

The variables $\omega_1, \omega_2$, and $\omega_3$ are weights that reflect the preference over the two stages' performances and are selected by the decision makers. However, these three variables are exogenous variables that cannot be determined by the two-stage model. In this study, we set $\omega_1 = \omega_2 = \omega_3 = 1$ because both operational efficiency and R&D efficiency are equal in importance to market efficiency in the R&D regions.

The variables $\lambda_j$ and $\mu_j$ represent the weight of the jth region in the first stage, while $\varphi_j$ is the weight in the second stage.

$\theta^O$ and $\theta^R$ represent the efficiency scores of operations and R&D in the first stage. $\theta^M$ is the efficiency score in the second stage. $\tilde{Z}_{dj}^O$ and $\tilde{Z}_{dj}^R$ represent unknown decision variables in the operations and R&D sectors in the intermediate measures. If $\theta^O = \theta^R = \theta^M = 1$ and the two-stages process is viewed as a whole; then, the value chain achieves an efficient performance. If $\theta^O = \theta^R = 1$ and $\theta^M > 1$ or ($\theta^O < 1, \theta^R < 1$ and $\theta^M = 1$), then model (8) indicates that one of the stages can achieve efficiency given a set of optimized intermediate measures.

As it is known, efficient DMU is not necessary to be MPSS. Thus, the efficiencies reported in Table 4 cannot give the policymakers accurate information on the scale size of the evaluated DMUs. Therefore, it is essential to know the scale size of the evaluated Chinese regions and select those regions that achieve the most productive scale size. In the following section, we introduce our MPSS model for the R&D value chain network of Chinese regions.

### 3.3 Specification of the R&D value chain MPSS model

In our proposed model, the operational, R&D, and market MPSSs of China's regional R&D activities are measured in a single DEA implementation. Consequently, through the new model, we determine the appropriate levels of sales and patents to achieve the most productive scale size.

The R&D value chain MPSS model for the Chin's provinces can be expressed as follows:

$$Chain's\ MPSS\ = \text{Max}\ \omega_1 \theta^M - \omega_2 \theta^O - \omega_3 \theta^R \tag{9}$$

$Stage\ 1: Operation$

$$s.t. \sum_{j=1}^{n} \lambda_j X_{ij}^O \leq \theta^O X_{io}^O, i = 1,2,\dots,m$$

$$\sum_{j=1}^{n} \lambda_j Z_{dj}^O \geq \tilde{Z}_{do}^O, d = 1,2,\dots,p$$

$$\sum_{j=1}^{n} \lambda_j = 1$$

$$\lambda_j, \theta^O \geq 0, j = 1,2,\dots,n$$

$Stage\ 1: R\&D$

$$\sum_{j=1}^{n} \mu_j X_{kj}^R \leq \theta^R X_{ko}^R, k = 1,2,\dots,K$$

$$\sum_{j=1}^{n} \mu_j Z_{ej}^R \geq \tilde{Z}_{eo}^R, e = 1,2,\ldots,E$$

$$\sum_{j=1}^{n} \mu_j = 1$$

$$\mu_j, \theta^R \geq 0, j = 1,2,\ldots,n$$

Stage 2 : Market

$$\sum_{j=1}^{n} \varphi_j Z_{dj}^O \leq \tilde{Z}_{do}^O, d = 1,2,\ldots,p$$

$$\sum_{j=1}^{n} \varphi_j Z_{ej}^R \leq \tilde{Z}_{eo}^R, e = 1,2,\ldots,E$$

$$\sum_{j=1}^{n} \varphi_j Y_{rj} \geq \theta^M Y_{ro}, r = 1,2,\ldots,s$$

$$\sum_{j=1}^{n} \varphi_j = 1$$

$$\varphi_j, \theta^M \geq 0, j = 1,2,\ldots,n.$$

Some explanations on models (8) and (9) are required. Model (8) restricted the distance measures $\theta^O$ and $\theta^R$ to be less than or equal one, and $\theta^M$ to be more than or equal one as the $\theta^O$ and $\theta^R$ are the input-oriented efficiencies of the operation and R&D processes and $\theta^M$ is the output-oriented efficiency of the market stage. Thus, the objective function value of model (8) is always less than or equal to one. The region is overall efficient if the objective function value is one. In contrast, model (9) relaxed the previous constraints, the distance measures $\theta^O$, $\theta^R$ and $\theta^M$, to be non-negative. Assuming $\theta^M \geq \theta^O + \theta^R$, the objective function value of model (9) is always non-negative. Based on MPSS definition, the region is overall MPSS if the objective function value of model (9) is zero.

The variables $\omega_1, \omega_2$, and $\omega_3$ are weights that reflect the decision makers' preference over the two stages' performances. Similarly, here, these weights cannot be determined by the MPSS model. In MPSS calculation, we set $\omega_1 = \omega_2 = \omega_3 = 1$ because both operational efficiency and R&D efficiency are equal in importance to market efficiency in the R&D regions.

This model not only calculates the MPSS of the evaluated regions based on the interrelationships of the internal processes but also estimates the appropriate values for a series of value-added production-related activities in which the two stages represent MPSS. Specifically, through the proposed MPSS model, set of unknown decision variables, $\tilde{Z}_{dj}^O$ and $\tilde{Z}_{dj}^R$, that identify the target setting of the original intermediate measures are generated to link the profitability stage with the marketability stage.

Model (9) can measure the MPSS of the whole R&D value chain but cannot measure the MPSS of each stage. To measure the MPSS of each stage, we can depend on the MPSS decomposition of the series network structure described in (Assani et al., 2018). In this case, we only measure the MPSS for one stage. The MPSS model for the first stage is proposed as follows:

$$Profitability's\ MPSS = \text{Max}(\theta^2 - \theta^1) + (\theta^4 - \theta^3) \quad (10)$$

$Stage\ 1 : Operation$

$$s.t. \sum_{j=1}^{n} \lambda_j X_{ij}^O \leq \theta^1 X_{io}^O, i = 1,2,\dots,m$$

$$\sum_{j=1}^{n} \lambda_j Z_{dj}^O \geq \theta^2 Z_{do}^O, d = 1,2,\dots,p$$

$$\sum_{j=1}^{n} \lambda_j = 1$$

$$\lambda_j, \theta^1, \theta^2 \geq 0, j = 1,2,\dots,n$$

$Stage\ 1 : R\&D$

$$\sum_{j=1}^{n} \mu_j X_{kj}^R \leq \theta^3 X_{ko}^R, k = 1,2,\dots,K$$

$$\sum_{j=1}^{n} \mu_j Z_{ej}^R \geq \theta^4 Z_{eo}^R, e = 1,2,\dots,E$$

$$\sum_{j=1}^{n} \mu_j = 1$$

$$\mu_j, \theta^3, \theta^4 \geq 0, j = 1,2,\dots,n$$

$Stage\ 2 : Market$

$$\sum_{j=1}^{n} \varphi_j Z_{dj}^O \leq \theta^2 Z_{do}^O, d = 1,2,\dots,p$$

$$\sum_{j=1}^{n} \varphi_j Z_{ej}^R \leq \theta^4 Z_{eo}^R, e = 1,2,\dots,E$$

$$\sum_{j=1}^{n} \varphi_j Y_{rj} \geq \theta^M Y_{ro}, r = 1,2,\ldots,s$$

$$\sum_{j=1}^{n} \varphi_j = 1$$

$$\varphi_j, \theta^M \geq 0, j = 1,2,\ldots,n$$

$$Chain's\ MPSS^* \ = \theta^M - \theta^1 - \theta^3.$$

In model (9), $\theta^1$ and $\theta^2$ are scalars representing expansion or contraction factors applied to the inputs and outputs of the operational process. $\theta^3$ and $\theta^4$ are scalars representing expansion or contraction factors applied to the inputs and outputs of the R&D process. The intermediate measures are adjusted radially as the inputs and outputs. The MPSS of the chain is kept unchanged in model (9) can be used to measure the MPSS of the whole system and to generate the appropriate intermediate measures (sales and patents), while model (10) can only measure the MPSS for each stage.

## 4 Empirical results and analysis

### 4.1 Data sources

Inputs and outputs data of chosen sectors were collected from the National Bureau of Statistics of China. Table 2 provides the descriptive statistics of inputs/outputs for China's regional R&D activities.

**Table 1 Summary of inputs and outputs descriptive statistics of China's regional R&D activities from 2014 to 2015**

|  | Mean | S.D. | Minimum | Maximum |
|---|---|---|---|---|
| **2015** | | | | |
| Employees | 1,324.45 | 1,127.2 | 121 | 5,935 |
| Investment in fixed assets (100 million Yuan) | 15,329 | 20,023 | 1,071 | 102,657 |
| R&D personnel | 85,106.1 | 113,831 | 43 | 441,304 |
| R&D projects | 9,996.6 | 13,868.7 | 21 | 51,940 |
| R&D expenditure (10000 Yuan) | 3,230,301 | 4,139,459 | 2,602 | 15,205,497 |
| Sales volumes (100 million Yuan) | 35,613.8 | 38,377.4 | 126.1 | 147,392 |
| Number of patents | 30,880.3 | 37,059.7 | 128 | 154,608 |
| Market value (100 million Yuan) | 30,220.4 | 63,409.7 | 0.1 | 345,389 |
| **2014** | | | | |
| Employees | 1,316.4 | 1,115.2 | 112 | 5,980 |
| Investment in fixed assets (100 million Yuan) | 13,598.9 | 15,998.4 | 688 | 69,113 |
| R&D personnel | 85,212.3 | 110,762.6 | 130 | 424,872 |
| R&D projects | 11,048.6 | 13,888 | 30 | 53,117 |
| R&D expenditure (10000 Yuan) | 2,985,245 | 3,749,636 | 2,943 | 13,765,378 |
| Sales volumes (100 million Yuan) | 35,232.2 | 36,782.8 | 109.3 | 141,194 |
| Number of patents | 25,474.1 | 32,077.7 | 92 | 146,660 |
| Market value (100 million Yuan) | 26,043.1 | 56,905.7 | 0.1 | 313,719 |

The descriptive statistics of the input and output variables in the value chain DEA model for 2014 and 2015 are presented in Table 2. The mean number of employees was 1,316.4 in 2014 and 1,324.45 in 2015. The mean of investments in the fixed assets were approximately 13,598.9 in 2014 and 15,329 hundred million yuan in 2015. The average R&D expenditures were 2,985,245 ten thousand and 3,230,301 ten thousand in 2014 and 2015, respectively. In addition, the mean number of patents was 25,474.1 in 2014 and 30,880.3 in 2015. The standard deviation of patents was 32,077.7 and 37,059.7 in the samples for 2014 and 2015, respectively.

## 4.2 Analysis of profitability and marketability efficiencies

The original efficiency values from 2014 to 2015 are presented in Table 4. The descriptive statistics of the regions' profitability and marketability efficiencies are presented in Table 3. For profitability efficiency, there are two different types of efficiency for each region: operational and R&D efficiency.

**Table 2 Summary statistics for the efficiency scores of China's R&D regions**

| Efficiency | Mean | S.D. | Minimum |
|---|---|---|---|
| **2014** | | | |
| Operation efficiency | 0.377 | 0.231 | 0.145 |
| R&D efficiency | 0.845 | 0.263 | 0.169 |
| Marketability efficiency | 0.119 | 0.201 | 0.004 |
| **2015** | | | |
| Operation efficiency | 0.508 | 0.306 | 0.140 |
| R&D efficiency | 0.833 | 0.268 | 0.162 |
| Marketability efficiency | 0.126 | 0.207 | 0.006 |

The average operational efficiency scores in 2014 and 2015 were 0.377 and 0.508, respectively, and the standard deviations were 0.231 and 0.306, respectively. In the original efficiency values, only two regions (Regions 29 and 30) and four regions (Regions 21, 24, 25, and 30) reached 100% efficiency during the production stage in 2014 and 2015.

The average values of R&D efficiency were 0.845 and 0.833 in 2014 and 2015, respectively, while the standard deviations were 0.263 and 0.268, respectively. In terms of R&D efficiency, there are 19 regions (Regions 1, 3, 4, 5, 6, 7, 8, 14, 20, 21, 22, 23, 24, 25, 27, 28, 29, 30, and 31) and 19 regions (Regions 1, 4, 5, 6, 7, 8, 14, 18, 20, 21, 22, 23, 24, 25, 27, 28, 29, 30, and 31) that attained appropriate efficiency levels for the initial efficiency scores in 2014 and 2015, but 18 regions (Regions 1, 4, 5, 6, 7, 8, 14, 20, 21, 22, 23, 24, 25, 27, 28, 29, 30, and 31) had consistent R&D efficiency levels in 2014–2015. Therefore, the average R&D efficiency was larger than the operational efficiency in 2014 and 2015. These results imply that the high-technology industry places emphasis on research and development activities rather than traditional production activities.

In terms of marketability efficiency, the average efficiency values were 0.119 and 0.126, and the standard deviations were 0.201 and 0.207 in 2014 and 2015, respectively. Region 1 (Beijing) has reached 100% marketability efficiency in 2014 and 2015. The top five regions that attained the highest marketability efficiency score are regions 1, 29, 27, 9, and 17 and regions 1, 29, 27, 17, and 9 in 2014 and 2015, respectively.

Based on the above analysis, the marketability efficiency values were low. As a result, a majority of the large high-technology firms in the Chinese regions performed inefficiently in terms of both profitability and marketability. This finding provides initial evidence that the generally lower profitability and marketability efficiency of high-technology firms in the Chinese regions is a serious problem that may be due to wasted resources on production and R&D. Interestingly, only two regions (Region 1 and 29) had appropriate efficiency levels in operations, R&D, and marketability efficiency.

Table 3 The efficiency scores and ranking of China's regional R&D value chain from 2014 to 2015

| Region | Operation efficiency | | | | R&D efficiency | | | | Marketability efficiency | | | |
|---|---|---|---|---|---|---|---|---|---|---|---|---|
| | 2014 | Rank | 2015 | R | 2014 | Rank | 2015 | R | 2014 | Rank | 2015 | R |
| 1 | 0.757 | 3 | 0.979 | 5 | 1.000 | 1 | 1.000 | 1 | 1.000 | 1 | 1.000 | 1 |
| 2 | 0.331 | 12 | 0.363 | 17 | 0.731 | 24 | 0.692 | 24 | 0.123 | 9 | 0.145 | 9 |
| 3 | 0.243 | 21 | 0.259 | 24 | 1.000 | 1 | 0.903 | 20 | 0.009 | 29 | 0.011 | 28 |
| 4 | 0.281 | 17 | 0.602 | 12 | 1.000 | 1 | 1.000 | 1 | 0.051 | 17 | 0.044 | 16 |
| 5 | 0.281 | 16 | 0.642 | 10 | 1.000 | 1 | 1.000 | 1 | 0.017 | 23 | 0.016 | 24 |
| 6 | 0.204 | 26 | 0.237 | 25 | 1.000 | 1 | 1.000 | 1 | 0.072 | 13 | 0.105 | 10 |
| 7 | 0.222 | 24 | 0.534 | 13 | 1.000 | 1 | 1.000 | 1 | 0.036 | 18 | 0.025 | 21 |
| 8 | 0.195 | 27 | 0.391 | 16 | 1.000 | 1 | 1.000 | 1 | 0.093 | 10 | 0.091 | 12 |
| 9 | 0.470 | 9 | 0.612 | 11 | 0.651 | 26 | 0.681 | 25 | 0.188 | 4 | 0.192 | 5 |
| 10 | 0.150 | 30 | 0.151 | 29 | 0.169 | 31 | 0.162 | 31 | 0.173 | 7 | 0.165 | 8 |
| 11 | 0.292 | 14 | 0.306 | 20 | 0.304 | 28 | 0.286 | 28 | 0.027 | 22 | 0.028 | 20 |
| 12 | 0.344 | 11 | 0.355 | 18 | 0.820 | 23 | 0.757 | 22 | 0.054 | 16 | 0.055 | 14 |
| 13 | 0.521 | 8 | 0.478 | 14 | 0.822 | 22 | 0.703 | 23 | 0.012 | 27 | 0.015 | 25 |
| 14 | 0.436 | 10 | 0.665 | 9 | 1.000 | 1 | 1.000 | 1 | 0.033 | 20 | 0.030 | 19 |
| 15 | 0.176 | 28 | 0.174 | 28 | 0.262 | 29 | 0.245 | 29 | 0.079 | 11 | 0.089 | 13 |
| 16 | 0.244 | 20 | 0.233 | 26 | 0.713 | 25 | 0.661 | 26 | 0.013 | 25 | 0.013 | 26 |
| 17 | 0.291 | 15 | 0.268 | 22 | 0.905 | 21 | 0.873 | 21 | 0.185 | 5 | 0.228 | 4 |
| 18 | 0.274 | 18 | 0.291 | 21 | 0.959 | 20 | 1.000 | 1 | 0.031 | 21 | 0.034 | 18 |
| 19 | 0.161 | 29 | 0.140 | 30 | 0.209 | 30 | 0.202 | 30 | 0.131 | 8 | 0.191 | 6 |
| 20 | 0.219 | 25 | 0.457 | 15 | 1.000 | 1 | 1.000 | 1 | 0.010 | 28 | 0.005 | 31 |
| 21 | 0.630 | 5 | 1.000 | 1 | 1.000 | 1 | 1.000 | 1 | 0.004 | 31 | 0.012 | 27 |
| 22 | 0.644 | 4 | 0.880 | 7 | 1.000 | 1 | 1.000 | 1 | 0.069 | 14 | 0.020 | 23 |
| 23 | 0.266 | 19 | 0.199 | 27 | 1.000 | 1 | 1.000 | 1 | 0.075 | 12 | 0.092 | 11 |
| 24 | 0.572 | 6 | 1.000 | 1 | 1.000 | 1 | 1.000 | 1 | 0.036 | 19 | 0.035 | 17 |
| 25 | 0.299 | 13 | 1.000 | 1 | 1.000 | 1 | 1.000 | 1 | 0.068 | 15 | 0.052 | 15 |
| 26 | 0.235 | 23 | 0.342 | 19 | 0.641 | 27 | 0.654 | 27 | 0.015 | 24 | 0.020 | 22 |
| 27 | 0.144 | 31 | 0.140 | 31 | 1.000 | 1 | 1.000 | 1 | 0.296 | 3 | 0.376 | 3 |
| 28 | 0.236 | 22 | 0.267 | 23 | 1.000 | 1 | 1.000 | 1 | 0.183 | 6 | 0.170 | 7 |
| 29 | 1.000 | 1 | 0.952 | 6 | 1.000 | 1 | 1.000 | 1 | 0.582 | 2 | 0.619 | 2 |
| 30 | 1.000 | 1 | 1.000 | 1 | 1.000 | 1 | 1.000 | 1 | 0.012 | 26 | 0.011 | 29 |
| 31 | 0.566 | 7 | 0.833 | 8 | 1.000 | 1 | 1.000 | 1 | 0.009 | 30 | 0.006 | 30 |

## 4.3  Analysis of profitability and marketability MPSSs

The original MPSS values from 2014 to 2015 are presented in Table 5. For profitability MPSS, there are two different types of MPSS for each region: operational and R&D MPSS. In the original MPSS values, only four regions (Regions 11, 13, 14, and 15) and five regions (Regions 9, 10, 11, 13, and 15) are MPSS during the production stage in 2014 and 2015. Regions 11, 13, and 15 are MPSS in both years 2014-2015.

**Table 4 The MPSS score and rank of the profitability stage of the R&D value chain**

| | Operation MPSS | | | | R&D MPSS | | | | Profitability MPSS | | | |
|---|---|---|---|---|---|---|---|---|---|---|---|---|
| N0 | 2014 | R | 2015 | R | 2014 | R | 2015 | R | 2014 | R | 2015 | R |
| 1 | 4.960 | 23 | 4.282 | 22 | 0.000 | 1 | 0.000 | 1 | 4.960 | 14 | 4.282 | 13 |
| 2 | 2.314 | 18 | 1.818 | 14 | 2.609 | 12 | 2.427 | 13 | 4.923 | 13 | 4.245 | 12 |
| 3 | 0.907 | 12 | 0.910 | 11 | 11.506 | 23 | 7.544 | 21 | 12.414 | 20 | 8.454 | 18 |
| 4 | 3.500 | 22 | 6.047 | 24 | 9.488 | 20 | 12.272 | 25 | 12.989 | 21 | 18.319 | 24 |
| 5 | 3.357 | 21 | 2.920 | 19 | 52.775 | 28 | 39.875 | 30 | 56.132 | 28 | 42.796 | 28 |
| 6 | 1.110 | 13 | 2.393 | 16 | 3.196 | 14 | 3.207 | 14 | 4.305 | 12 | 5.600 | 16 |
| 7 | 2.352 | 19 | 2.414 | 17 | 10.795 | 22 | 10.700 | 23 | 13.147 | 22 | 13.114 | 22 |
| 8 | 6.569 | 26 | 8.532 | 26 | 3.358 | 16 | 3.292 | 16 | 9.927 | 19 | 11.824 | 20 |
| 9 | 0.570 | 9 | 0.000 | 1 | 1.345 | 8 | 1.212 | 10 | 1.915 | 5 | 1.212 | 5 |
| 10 | 0.015 | 6 | 0.000 | 1 | 0.363 | 2 | 0.413 | 4 | 0.379 | 1 | 0.413 | 1 |
| 11 | 0.000 | 1 | 0.000 | 1 | 1.187 | 7 | 1.028 | 8 | 1.187 | 4 | 1.028 | 4 |
| 12 | 1.327 | 15 | 1.442 | 12 | 0.744 | 3 | 0.544 | 5 | 2.070 | 6 | 1.986 | 7 |
| 13 | 0.000 | 1 | 0.000 | 1 | 6.854 | 19 | 4.334 | 19 | 6.854 | 17 | 4.334 | 14 |
| 14 | 0.000 | 1 | 0.065 | 6 | 16.603 | 25 | 13.882 | 26 | 16.603 | 23 | 13.948 | 23 |
| 15 | 0.000 | 1 | 0.000 | 1 | 0.748 | 4 | 0.706 | 7 | 0.748 | 2 | 0.706 | 2 |
| 16 | 0.071 | 8 | 0.072 | 8 | 3.264 | 15 | 3.506 | 18 | 3.335 | 9 | 3.578 | 11 |
| 17 | 0.749 | 10 | 0.801 | 10 | 2.562 | 11 | 2.071 | 12 | 3.311 | 8 | 2.872 | 9 |
| 18 | 1.777 | 16 | 2.134 | 15 | 4.478 | 18 | 3.424 | 17 | 6.255 | 16 | 5.558 | 15 |
| 19 | 0.003 | 5 | 0.068 | 7 | 0.830 | 6 | 0.653 | 6 | 0.833 | 3 | 0.721 | 3 |
| 20 | 2.955 | 20 | 2.927 | 20 | 0.750 | 5 | 0.000 | 1 | 3.705 | 11 | 2.927 | 10 |
| 21 | 46.894 | 30 | 45.906 | 30 | 59.594 | 30 | 51.611 | 31 | 106.488 | 30 | 97.517 | 30 |
| 22 | 0.771 | 11 | 0.097 | 9 | 2.621 | 13 | 1.312 | 11 | 3.392 | 10 | 1.409 | 6 |
| 23 | 1.128 | 14 | 1.525 | 13 | 1.419 | 9 | 1.056 | 9 | 2.547 | 7 | 2.581 | 8 |
| 24 | 1.975 | 17 | 3.308 | 21 | 3.831 | 17 | 6.460 | 20 | 5.806 | 15 | 9.768 | 19 |
| 25 | 6.483 | 25 | 2.655 | 18 | 11.991 | 24 | 10.174 | 22 | 18.474 | 24 | 12.829 | 21 |
| 26 | 1238.33 | 31 | 1113.36 | 31 | 55.817 | 29 | 0.000 | 1 | 1294.14 | 31 | 1113.36 | 31 |
| 27 | 5.304 | 24 | 4.656 | 23 | 1.749 | 10 | 3.271 | 15 | 7.053 | 18 | 7.926 | 17 |
| 28 | 11.380 | 28 | 13.310 | 28 | 10.642 | 21 | 11.136 | 24 | 22.022 | 25 | 24.446 | 25 |
| 29 | 23.047 | 29 | 31.931 | 29 | 60.621 | 31 | 30.266 | 29 | 83.668 | 29 | 62.196 | 29 |
| 30 | 8.987 | 27 | 11.826 | 27 | 23.271 | 27 | 21.688 | 28 | 32.258 | 27 | 33.514 | 27 |
| 31 | 0.032 | 7 | 7.987 | 25 | 23.061 | 26 | 21.636 | 27 | 23.093 | 26 | 29.623 | 26 |

In terms of R&D MPSS, there is one region (Region 1) and three regions (Regions 1, 20, and 26) that are MPSS in 2014 and 2015, but only one region (Region 1) is MPSS in both years. As it is shown in Table 5, no regions are MPSS in the profitability stage. It is clear that the profitability MPSS is the sum of the operation and R&D efforts MPSSs obeying the MPSS decomposition of the parallel network described in (Assani et al., 2019).

In terms of marketability MPSS, only Region 1 (Beijing) was MPSS in both years (see Table 6). That is, the production and R&D efforts did not sufficiently reflect the regions' market valuations. As a result, most of these regions performed inefficiently in terms of both profitability and marketability. This finding provides initial evidence that the generally lower profitability and marketability efficiency of Chinese R&D regions is a severe problem that may be due to wasted resources on production and R&D. Interestingly, only three regions (Regions 10, 15, and 19) had appropriate MPSS levels in operations, R&D, and marketability efficiency. Therefore, the various intermediate resource inputs and outcomes must determine the level of effort necessary to boost overall productivity. This problem is particularly impressive given that nearly all previous studies have ignored these intermediate measures in R&D activities.

**Table 5 The MPSS score and rank of the marketability stage and of the R&D value chain**

| | Marketability MPSS | | | | R&D value chain MPSS | | | |
|---|---|---|---|---|---|---|---|---|
| Region | 2014 | Rank | 2015 | R | 2014 | Rank | 2015 | R |
| 1 | 0.000 | 1 | 0.000 | 1 | 4.960 | 8 | 4.282 | 9 |
| 2 | 2.430 | 13 | 2.472 | 12 | 7.353 | 14 | 6.717 | 14 |

| | | | | | | | |
|---|---|---|---|---|---|---|---|
| 3 | 3.131 | 18 | 2.367 | 11 | 9.283 | 16 | 10.820 | 18 |
| 4 | 2.907 | 16 | 3.095 | 16 | 15.895 | 21 | 21.414 | 24 |
| 5 | 12.034 | 25 | 14.247 | 23 | 68.166 | 28 | 57.043 | 28 |
| 6 | 1.504 | 8 | 1.654 | 10 | 5.810 | 11 | 7.254 | 15 |
| 7 | 2.750 | 15 | 9.286 | 21 | 10.398 | 18 | 3.828 | 7 |
| 8 | 1.781 | 10 | 2.650 | 13 | 11.708 | 20 | 14.474 | 20 |
| 9 | 3.071 | 17 | 1.282 | 7 | 4.986 | 9 | 2.494 | 5 |
| 10 | 0.155 | 2 | 0.468 | 6 | 0.534 | 1 | 0.881 | 3 |
| 11 | 5.954 | 22 | 6.296 | 17 | 4.768 | 7 | 5.268 | 10 |
| 12 | 1.623 | 9 | 1.595 | 9 | 3.693 | 6 | 3.581 | 6 |
| 13 | 1.251 | 7 | 25.047 | 26 | 5.604 | 10 | 20.713 | 23 |
| 14 | 11.067 | 24 | 15.688 | 24 | 27.670 | 24 | 1.741 | 4 |
| 15 | 0.398 | 4 | 0.071 | 2 | 1.146 | 3 | 0.777 | 1 |
| 16 | 5.101 | 21 | 8.907 | 20 | 1.767 | 4 | 5.328 | 11 |
| 17 | 2.597 | 14 | 2.663 | 14 | 5.908 | 12 | 5.535 | 12 |
| 18 | 4.377 | 19 | 0.198 | 4 | 10.632 | 19 | 5.756 | 13 |
| 19 | 0.211 | 3 | 0.136 | 3 | 1.043 | 2 | 0.858 | 2 |
| 20 | 23.081 | 26 | 10.639 | 22 | 19.376 | 22 | 7.712 | 16 |
| 21 | 388.092 | 30 | 116.967 | 31 | 281.604 | 30 | 19.450 | 22 |
| 22 | 4.982 | 20 | 23.633 | 25 | 8.374 | 15 | 22.224 | 25 |
| 23 | 0.684 | 6 | 1.423 | 8 | 3.232 | 5 | 4.004 | 8 |
| 24 | 0.458 | 5 | 0.311 | 5 | 6.265 | 13 | 9.457 | 17 |
| 25 | 2.090 | 11 | 6.600 | 18 | 20.564 | 23 | 19.429 | 21 |
| 26 | 13371.934 | 31 | 55.924 | 27 | 12077.791 | 31 | 1169.284 | 31 |

| | | | | | | | |
|---|---|---|---|---|---|---|---|
| 27 | 2.303 | 12 | 3.038 | 15 | 9.356 | 17 | 10.964 | 19 |
| 28 | 6.199 | 23 | 7.041 | 19 | 28.221 | 25 | 31.487 | 26 |
| 29 | 49.475 | 27 | 59.540 | 28 | 133.143 | 29 | 121.736 | 30 |
| 30 | 65.522 | 28 | 76.994 | 29 | 33.264 | 26 | 43.480 | 27 |
| 31 | 71.715 | 29 | 109.097 | 30 | 48.622 | 27 | 79.473 | 29 |

In addition, the non-parametric Kruskal–Wallis test is used to determine whether the rankings of the MPSS scores differed across the different period groups (Table 7). The results indicate that there were no significant differences found among the rankings of the operations, R&D, and marketability MPSS scores for 2014 and 2015. That is, the rankings of the MPSS scores among these groups showed a high degree of consistency from 2014 to 2015.

**Table 6 The Kruskal-Wallis test of MPSS scores ranking for 2014 and 2015**

|  | Operational | R&D | Marketability |
|---|---|---|---|
| Chi-Square | 0.119 | 0.557 | 0.294 |
| *Df* | 1 | 1 | 1 |
| Asymp. Sig. | 0.730 | 0.456 | 0.558 |

## 4.4 The intermediate outputs of Stage 1

The above discussion indicates that each of the two stages represents a non-dominant performance that is given a set of optimized intermediate measures determined by the value chain MPSS model. This model provides not only the MPSS measurement but also values that indicate the appropriate degree of intermediate measures for the two stages. That is, we can obtain directions for achieving the appropriate level of efficiency for this R&D value chain. Consequently, we can estimate the appropriate intermediate impacts of production and R&D efforts on the regions' performance. The results of the appropriate intermediate measures under the value chain MPSS model are shown in Table 8.

**Table 7 The appropriate levels of intermediate measures of profitability stage for 2015**

| Regions | Current level | | Appropriate level | | Gap | | Improving strategy |
|---|---|---|---|---|---|---|---|
| | Sales | Patents | Sales | Patents | Sales | Patents | |
| 1 | 17,279 | 88,930 | 3604.23 | 88,930 | -13,675 | 0 | Sales↓ |
| 2 | 27,460 | 28,510 | 14417.74 | 24630.17 | -13,043 | -3,880 | Sales↓, Patents↓ |
| 3 | 45,407 | 11,259 | 13666.27 | 9262.706 | -31,741 | -1,996 | |
| 6 | 32,927 | 19,332 | 12417.94 | 14714.26 | -20,509 | -4,618 | |
| 7 | 22,529 | 6,154 | 9557.201 | 5888.075 | -12,972 | -266 | |
| 8 | 11,524 | 14,663 | 9854.652 | 9016.993 | -1,669 | -5,646 | |
| 9 | 31,214 | 46,976 | 12205.98 | 33416.11 | -19,008 | -13,560 | |
| 10 | 147,392 | 154,608 | 26272.14 | 32742.35 | -121,120 | -121,866 | |
| 11 | 64,279 | 67,674 | 15799.03 | 11800.63 | -48,480 | -55,873 | |
| 12 | 38,798 | 68,314 | 13208.93 | 12320.07 | -25,589 | -55,994 | |
| 13 | 40,216 | 17,663 | 11788.21 | 9105.039 | -28,428 | -8,558 | |
| 15 | 144,234 | 93,475 | 31463.24 | 30196.09 | -112,770 | -63,279 | |
| 16 | 73,367 | 21,338 | 20605.4 | 14738.55 | -52,762 | -6,599 | |
| 17 | 44,113 | 30,204 | 11134.44 | 26646.14 | -32,979 | -3,558 | |
| 18 | 36,232 | 19,499 | 12870.41 | 8767.787 | -23,361 | -10,731 | |
| 19 | 121,050 | 103,941 | 23589.34 | 33999.53 | -97,460 | -69,941 | |
| 20 | 21,412 | 30,815 | 11443.03 | 6230.643 | -9,969 | -24,584 | |
| 22 | 20,945 | 35,086 | 8074.055 | 7979.008 | -12,871 | -27,107 | |
| 23 | 39,213 | 40,437 | 17432.6 | 18680.7 | -21,781 | -21,756 | |
| 24 | 9,821 | 7,538 | 6290.621 | 5061.953 | -3,530 | -2,476 | |
| 30 | 3,604 | 2,626 | 3198.178 | 2585.815 | -406 | -40 | |
| 4 | 12,567 | 5,680 | 8625.576 | 6346.337 | -3,941 | 666 | Sales↓, Patents↑ |
| 5 | 18,702 | 2,254 | 5437.341 | 2373.755 | -13,265 | 120 | |
| 14 | 30,618 | 5,722 | 12804.81 | 10539.51 | -17,814 | 4,818 | |
| 25 | 9,668 | 6,301 | 6154.199 | 6619.61 | -3,514 | 319 | |
| 27 | 20,248 | 17,322 | 8778.824 | 21821.65 | -11,469 | 4,500 | |
| 31 | 7,945 | 3,024 | 6876.113 | 3777.211 | -1,069 | 753 | |
| 21 | 1,833 | 1,211 | 2876.971 | 1902.661 | 1,044 | 692 | Sales↑, Patents↑ |
| 26 | 126 | 128 | 1200.378 | 600.215 | 1,074.258 | 472.215 | |
| 28 | 6,942 | 5,504 | 11580.22 | 10934.21 | 4,638 | 5,430 | |

| | | | | | | |
|---|---|---|---|---|---|---|
| 29 | 2,359 | 1,103 | 3209.493 | 1794.087 | 851 | 691 |

The first double columns of Table 8 report the current levels of the intermediate measures of the profitability stage. The second double columns are the appropriate levels of the intermediate measures of the profitability stage. The third set of double columns represents the gap between the appropriate and current levels of the intermediate measures. The improvement strategy is given for each region in the last column.

In terms of the intermediate measures, only one region (Region 1) attained the appropriate intermediate of patents, while it should decrease the other intermediate measure (sale) of the operation process. Such a result supports the fact that the success of the R&D firms comes not only from their R&D efforts but also from the harmony among their operational and management activities.

## 5 Conclusion

This study proposes new models to measure the most productive scale size of systems that have a mixed structure of series or parallel structures. Two properties of mixed structures have been discussed and examined.

The first property deals with a general multi-stage network where exogenous inputs for each stage are supplied, and there are intermediate measures connect the internal stages, and final outputs from each stage are obtained. We proposed a new network MPSS model to measure the MPSS of such networks. To measure the MPSS of the internal stages, we decomposed the system MPSS into the internal stages' MPSSs by converting the mixed structure network into a series of parallel processes using dummy processes. The original network and the tandem network are equivalent. The tandem network has the ability to decompose the overall efficiency and MPSS into the internal stages. An application of 24 non-life insurance companies is used to show the applicability and the merits of the proposed methods in both measuring and decomposing MPSS.

The second property considers a real-life application of China's regional R&D activities for 2014 and 2015. We build the R&D value chain network as a mixed structure network composed of two stages, where the first stage has two processes connected in parallel. The first stage is the profitability stage, which has operational and R&D efforts processes connected in parallel. The operational process consumes the employees and the investment in the fixed assets as inputs and produces the sale volume as output. The second process is the R&D efforts that use the R&D personnel, R&D projects, and R&D expenditure as inputs and produce the patents as output. The outputs the profitability stage are intermediate measures used as inputs of the marketability stage to produce the market value as the final outputs. We measured the operational, R&D, and the marketability efficiencies for 2014 and 2015. Then we proposed MPSS models to measure the system and the stages' MPSSs. The MPSS network model provides not only the MPSS measurement but also values that indicate the appropriate degree of intermediate measures for the two stages. That is, we can obtain directions for achieving the appropriate level of efficiency for this R&D value chain. Consequently, we can estimate the appropriate intermediate impacts of production and R&D efforts on the regions' performance. Improvement's strategy is given for each Chinese region based on the gap between the current and the appropriate level of intermediate measures. Our findings show that the marketability efficiency values of Chinese R&D regions were low. As a result, a majority of the large high-technology firms in the Chinese regions performed inefficiently in terms of both profitability and marketability. This finding provides initial evidence that the generally lower profitability and marketability efficiency of high-technology firms in the Chinese regions is a severe problem that may be due to wasted resources on production and R&D. Interestingly, only two regions (Beijing and Qinghai) had appropriate levels in operations, R&D, and marketability efficiencies. In terms of MPSS, no regions were MPSS in the profitability stage, while Beijing was the only MPSS region in the marketability stage. Jiangsu, Shandong, and Guangdong had appropriate levels in operations, R&D, and marketability MPSSs. The results of the Kruskal–Wallis test indicate that there were no significant differences found among the rankings of the operations, R&D, and marketability MPSS scores for 2014 and 2015.